\documentclass[12pt,preprint]{emulateapj}

\newcommand{\SDSS}{SDSS\,J090221.35+381941.9}
\newcommand{\SDSSs}{SDSS\,J0902}
\def\msun{M$_{\odot}$}

\def\asec{\ifmmode ^{\prime\prime}\else$^{\prime\prime}$\fi}
\def\farcs{\hbox{$.\!\!^{\prime\prime}$}}  	
\def\degs{\ifmmode ^{\circ}\else$^{\circ}$\fi}

\shorttitle{Four New AM\,CVn Stars}
\shortauthors{Rau et al.}

\begin{document}


\title{A census of AM CVn stars: three new candidates and one confirmed 48.3-minute binary}


\author{A. Rau\altaffilmark{1,2}, G.~H.~A. Roelofs\altaffilmark{3},  P.~J. Groot\altaffilmark{4},  T.~R. Marsh\altaffilmark{5}, G. Nelemans\altaffilmark{4}, D. Steeghs\altaffilmark{3,5}, M.~M. Kasliwal\altaffilmark{2}}


\email{arau@mpe.mpg.de}

\altaffiltext{1}{Max-Planck Institute for Extraterrestrial Physics, Giessenbachstr, Garching, 85748, Germany}
\altaffiltext{2}{Caltech Optical Observatories, MS 105-24, California Institute of Technology, Pasadena, CA 91125, USA}
\altaffiltext{3}{Harvard-Smithsonian Center for Astrophysics, 60 Garden Street, Cambridge, MA 02138, USA}
\altaffiltext{4}{Department of Astrophysics, IMAPP, Radboud University Nijmegen, PO Box 9010, 6500 GL Nijmegen, The Netherlands}
\altaffiltext{5}{Department of Physics, University of Warwick, Coventry CV4 7Al, UK}
\begin{abstract}

We present three new candidate AM CVn binaries, plus one confirmed
new system, from a spectroscopic survey of color-selected objects
from the Sloan Digital Sky Survey. All four systems were found from
their helium emission lines in low-resolution spectra taken on the
Hale telescope at Palomar, and the Nordic Optical Telescope and the
William Herschel Telescope on La Palma.

The ultra-compact binary nature of \SDSS\ was confirmed using
phase-resolved spectroscopy at the Keck-I telescope.  From the
characteristic radial velocity `S-wave' observed in the helium
emission lines we measure an orbital period of
48.31$\pm$0.08\,min. The continuum emission can be described with a
blackbody or a helium white dwarf atmosphere of T$_{\rm
  eff}\sim15,000$\,K, in agreement with theoretical cooling models
for relatively massive accretors and/or donors.  The absence in the
spectrum of broad helium absorption lines from the accreting white
dwarf suggests that the accreting white dwarf cannot be much hotter
than 15,000\,K, or that an additional component such as the
accretion disk contributes substantially to the optical flux.

Two of the candidate systems, SDSS~J152509.57+360054.5 and
SDSS~J172102.48+273301.2, do show helium absorption in the blue part
of their spectra in addition to the characteristic helium emission
lines. This, in combination with the high effective temperatures of
$\sim18,000$\,K and $\sim16,000$\,K suggests both two be at orbital
periods below $\approx40$\,min.  The third candidate,
SDSS~J164228.06+193410.0, exhibits remarkably strong helium emission
on top of a relatively cool (T$_{\rm eff}\sim12,000$\,K) continuum,
indicating an orbital period above $\sim$50\,min.

\end{abstract}



\section{Introduction}
\label{sec:intro}

The  AM~CVn  stars  form   a  small  class  of  interacting  binaries,
consisting of white dwarf accretors with degenerate or semi-degenerate
helium-transferring   companion  stars.    AM  CVn   stars   exist  in
ultra-compact configurations,  with observed orbital  periods, $P_{\rm
  orb}$, ranging from 65\,min down to 10\,min \citep[below the orbital
period   minimum   for  hydrogen-rich   donors,   see][for  a   recent
review]{Nelemans:2005lr}.  Two  ultrashort-period candidates, V407~Vul
\citep{Cropper:1998aa}  and   HM~Cnc  \citep{Israel:2002aa},  possibly
extend the $P_{\rm orb}$-distribution further down to 5.3\,min.
 
AM CVn  stars are of interest  for binary stellar  evolution theory as
they   form  the   end-product  of   several   evolutionary  scenarios
\citep[e.g.,][]{Nelemans:2001aa,Podsiadlowski:2003aa}      and     can
potentially   produce  rare,   sub-luminous,  SN   Ia-like  explosions
\citep[SN .Ia,][]{Bildsten:2007xy}.  They are also the strongest known
sources that  can be detected as gravitational-wave  emitters with the
future {\it Laser Interferometer Space Antenna} \citep[{\it LISA}; see
e.g.,][]{Nelemans:2004aa,Stroeer:2005aa,Roelofs:2007ac}.

Our  understanding  of the  physics  of  these  systems has  increased
significantly        over       the        last        few       years
\citep[e.g.,][]{Deloye:2005aa,Bildsten:2006aa,Deloye:2007aa},     while
observationally,      the      Sloan      Digital      Sky      Survey
\citep[SDSS,][]{York:2000aa}  has  been  instrumental. Pre-SDSS,  only
$\sim$10 AM  CVn binaries  were known \citep{Roelofs:2005aa},  and the
extreme heterogeneity  of this already  small sample made a  census of
the Galactic population of AM  CVn stars impossible.  The SDSS-I (Data
Release 5) provided  a substantially more homogeneous sample  of 6 new
AM CVn stars, allowing for  a local space density measurement, limited
by the still small size of the sample \citep{Roelofs:2007ac}.

Noting that  the AM~CVn stars  occupy a relatively  sparsely populated
region  in color space  where the  spectroscopic completeness  down to
$g=20.5$  in the  SDSS database  is only  $\sim$20\,\%, we  started an
extensive spectroscopic  survey program  at a number  of observatories
world-wide to resolve  the `hidden' population of AM~CVn  stars in the
SDSS photometry.  The aim of this project is to significantly increase
the  total known  population of  AM~CVn stars  and,  more importantly,
enlarge the homogeneous sample of these sources. Ultimately, the space
density of  AM CVn stars, and further  population characteristics such
as  their orbital  period  distribution, will  allow comparisons  with
predictions from theoretical models, and will allow better modeling of
the Galactic gravitational-wave foreground signal seen by {\it LISA}.

Details  of our  spectroscopic  follow-up program,  in particular  the
sample selection  criteria, and the  first (peculiar) new AM  CVn star
(SDSS~J080449.49+161624.8) were described in \cite{Roelofs:2009aa}.  A
complete  catalog of sources,  and a detailed  AM CVn  population study,
will be  presented when the spectroscopic survey  program is complete.
Here, we  give a status update  after the first year  and present four
new  discoveries,  including a  detailed  spectroscopic  study of  one
confirmed new AM CVn star.

The  outline of  the paper  is  as follows.  In \S\,\ref{sec:data}  we
summarize the observations  and data reduction. We present  the new AM
CVn  candidates as well  as the  spectral and  timing analysis  of the
confirmed system in  \S\,\ref{sec:results}.  Our conclusions are given
in \S\,\ref{sec:conclusion}.


\section{Observations and Data reduction}
\label{sec:data}

\begin{figure*}[htb]
\begin{center}
\includegraphics[width=75mm]{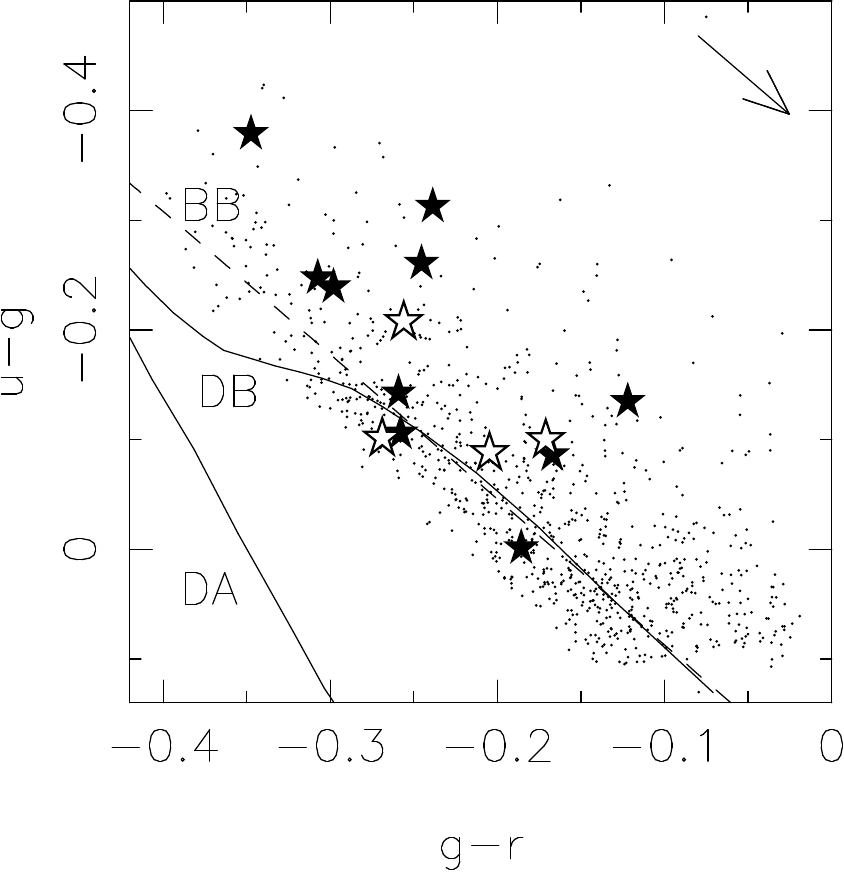}\hspace{0.0mm}%
\includegraphics[width=75mm]{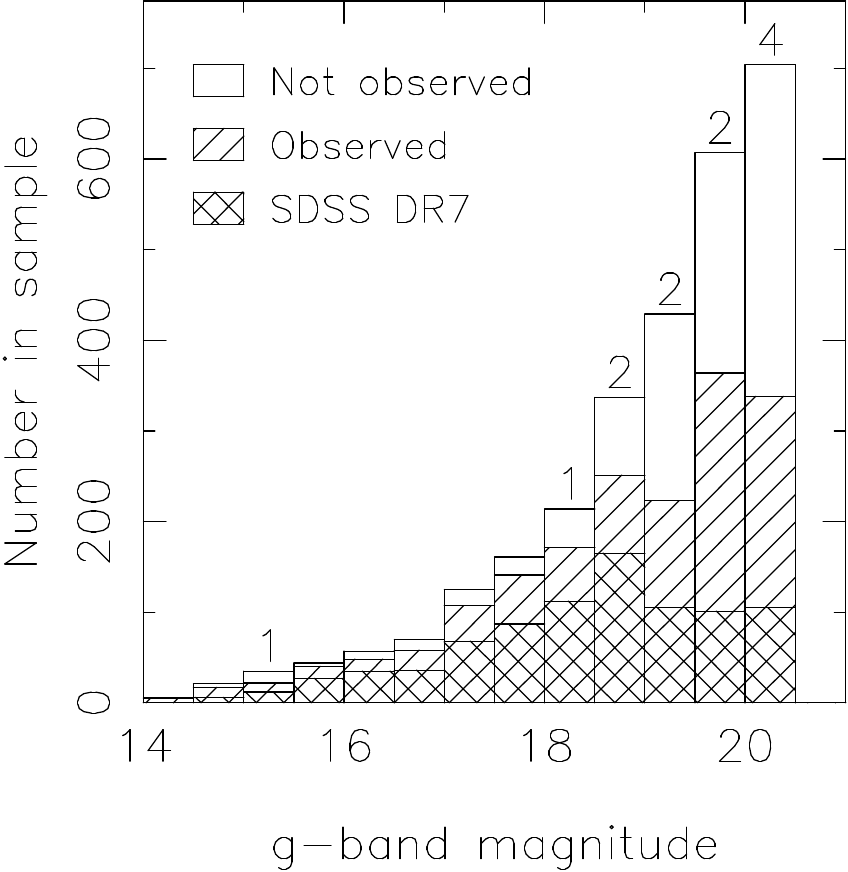}
\caption{Photometric colors (left, de-reddened) and spectroscopic completeness (right, not de-reddened) of our sample. Left, stars indicate known AM CVn stars with SDSS photometry; open stars are the  four new systems presented here (SDSS J0902, J1525, J1721, J1642 from left to right). The dots denote our candidate population up to DR7. Blackbody (BB), helium-atmosphere (DB) and hydrogen-atmosphere (DA) white dwarf cooling curves are indicated. The arrow shows the reddening vector for a $g$-band extinction of 0.2\,mag. Right, numbers above bins give the number of AM CVns in our spectroscopy or the DR7 spectroscopic database. Two systems with DR7 photometry have not been observed spectroscopically by us or the SDSS. The brightest system is the well studied source GP~Com. }
\label{fig:sample}
\end{center}
\end{figure*}

Low-resolution,  low-signal-to-noise ratio spectra  of objects  in our
color-selected sample have been obtained on a multitude of telescopes.
To date,  almost half  of our sample  of $\approx$1500  candidates has
been completed  (see Fig.\ \ref{fig:sample}  for a concise  summary of
the  current   statistics).   Two   of  the  systems   presented  here
(SDSS\,J172102.48+273301.2,       hereafter      SDSS\,J1721,      and
SDSS\,J152509.57+360054.5,  hereafter SDSS\,J1525) were  identified as
helium-emission-line objects  from spectra  taken at the  2.6-m Nordic
Optical  Telescope   (NOT)  on  La  Palma,  Spain,   with  the  ALFOSC
spectrograph     fitted      with     the     grism      \#11     (see
Table~\ref{tab:obslog}). Slits  with 0\farcs9 and  1\farcs3 width were
used for SDSS~J1721  and SDSS~J1525, respectively, offering resultions
of  R$\sim 210$  and  R$\sim  145$ at  5200\,\AA.   The third  object,
SDSS\,J164228.06+193410.0 (hereafter SDSS\,J1642)  was identified in a
spectrum taken at the 4.2-m  William Herschel Telescope (WHT), also on
La Palma, equipped  with the new Auxiliary-port Camera  (ACAM) and the
400V holographic grism, providing  an effective resolution $R\sim 450$
at  6000\,\AA.   This set-up  was  also  used  to obtain  confirmation
spectra of  the two objects  discovered at the  NOT.  The NOT  and WHT
spectra  were corrected for  Galacitic foreground  extinction assuming
that the AM~CVn stars are on average located behind the Galactic Plane
dust screen \citep{Roelofs:2009aa}.

The identifying spectrum of \SDSS\ (hereafter \SDSSs) was taken at the
Hale  5-m   telescope  at   Palomar  Observatory  equipped   with  the
Double-Beam Spectrograph \citep[DBSP;][]{oke:1982aa}.  A 1\farcs5 slit
and the 300/3990 and 316/7500 gratings were used, achieving a spectral
resolution on  the blue  and red sides  of R$\sim540$  (4000\,\AA) and
R$\sim700$ (6000\,\AA), respectively.

\begin{table}
\begin{center}
\caption{Log of spectroscopic observations.
\label{tab:obslog}}
\begin{tabular}{l l c r}
\hline\hline
SDSS\,\ldots      & UT Date       & Set-up        & Exposures \\
\hline
J090221.35+381941.9~ & 2009-01-01   & P200/DBSP     & $1\times910$\,s\\
            & 2009-01-26   & Keck-I/LRIS     & $20\times240$\,s\\
            & 2009-01-27   &               & \tablenotemark{$\dagger$}$45\times180$\,s\\[1ex]
J152509.57+360054.5~  & 2009-05-26   & NOT/ALFOSC    & $1\times600$\,s\\
            & 2009-06-25   & WHT/ACAM      & $1\times900$\,s\\[1ex]
J164228.06+193410.0~ & 2009-06-23   & WHT/ACAM      & $2\times400$\,s\\[1ex]
J172102.48+273301.2~ & 2009-05-25   & NOT/ALFOSC    & $1\times600$\,s\\
            & 2009-06-19   & WHT/ACAM      & $1\times600$\,s\\
\hline
\end{tabular}
\tablenotetext{$\dagger$}{43 exposures on the red side.}
\end{center}
\end{table}


Follow-up  phase-resolved  spectroscopy  of   \SDSSs\  was
obtained on  2009 Jan  26 and 27  using the Low-Resolution  Imager and
Spectrograph   \citep[LRIS;][]{Oke:1995aa}  at   the   10\,m  {Keck-I}
telescope (see Table~\ref{tab:obslog}).  Similar  to DBSP, LRIS is a
dual-beam instrument. We used  the 600/4000 grism and 600/7500 grating
for the  blue and red  arms, respectively, with  the light split  by a
dichroic at $\approx5600$\,\AA.  The blue  arm data were binned on the
chip by  factors of 4 each  in spatial and  dispersion direction. This
allowed  a reduction of  the overhead  due to  CCD readout  from 42\,s
(unbinned) to 27\,s.  All LRIS observations were performed with a long
slit  of 1\farcs5 width  and the  atmospheric dispersion  corrector in
place.  Typical  seeing was about  1\farcs2 resulting in  an effective
spectral  resolution   of  about  4.5\,\AA\  FWHM   at  4500\,\AA,  or
300\,km\,s$^{-1}$.  The red arm  data were binned $2\times2$ providing
readout times of  34\,s (compared to 72\,s unbinned)  and a resolution
of 4.4\,\AA\ at 6000\,\AA\ ($\approx220$\,km\,s$^{-1}$).

All data  were reduced with  standard IRAF routines, and  spectra were
extracted    using     an    optimal    (variance-weighted)    method.
Spectro-photometric flux  calibration was carried out  with spectra of
the   standard    star   Feige~34   \citep{Oke:1990aa}    taken   each
night.    Correction   for    Galactic   foreground    extinction   of
E(B--V)=0.024\,mag \citep{Schlegel:1998aa} was applied.


\section{Results}
\label{sec:results}

\subsection{Identification Spectra of the New AM CVns}

\begin{figure}[t]
\begin{center}
\includegraphics[width=0.49\textwidth,angle=0]{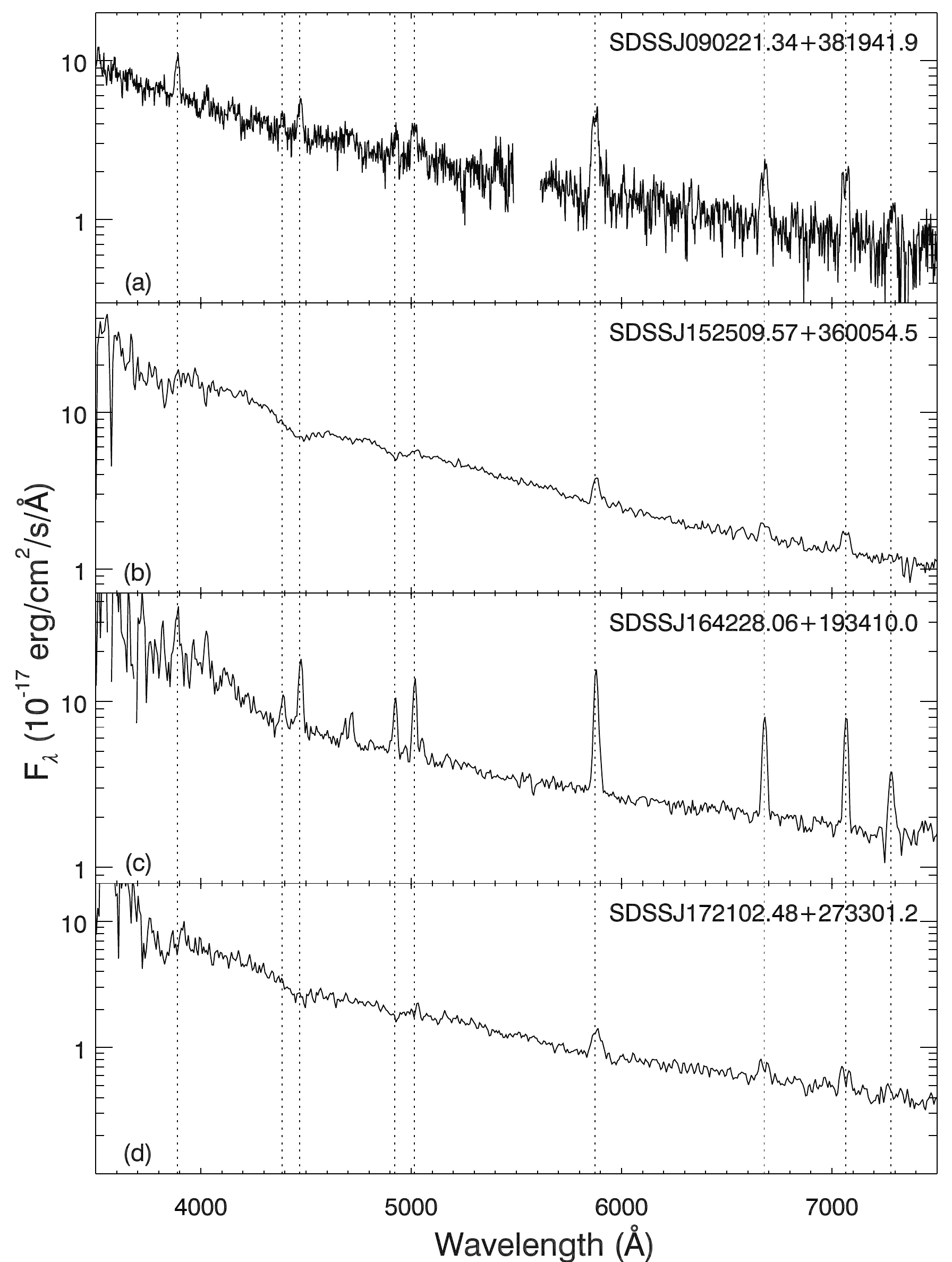}
\caption{De-reddened low-resolution spectra of (a) \SDSSs\ (taken with P200/DBSP), (b) SDSSJ~1525, (c) SDSSJ~1642, and (d) SDSSJ~1721 (all taken with WHT/ACAM). The positions of the strongest HeI  lines are indicated  by the dotted vertical lines.}
\label{fig:spectra}
\end{center}
\end{figure}

The   extracted   spectra  of   all   four   objects   are  shown   in
Figure~\ref{fig:spectra}.  The prominent neutral He emission lines are
clearly recognized in all of them, most notably \ion{He}{1} 5876, 6678
\& 7065.   This, and the absence  of hydrogen in  the spectra suggests
that all four  objects are likely new members of  the class of AM\,CVn
stars,  although  further observations  are  needed  to confirm  their
ultra-compact binary nature beyond question.

\subsection{Phase-Resolved Spectroscopy of  \SDSSs}

Phase-resolved spectroscopy  is the technique of  choice for measuring
the  orbital period  of AM  CVn  binaries, for  which photometry  does
typically not yield  stable orbital period signals.  In  AM CVn stars,
the  emission  lines  in  a  spectroscopic time  series  are  commonly
observed to  vary in the form  of a sinusoidal  ``S-wave''. In analogy
with  the bright  spots seen  in hydrogen-rich  cataclysmic variables,
this  S-wave  is caused  by  the  impact  region where  the  infalling
accretion  stream   hits  the   accretion  disk  edge,   providing  an
emission-line source  that is fixed in  the binary frame.  In order to
search for a similar signature in  our data of \SDSSs, we combined the
strongest helium lines from both nights of {Keck-I/LRIS} observations.
We  focused on  the blue  side of  LRIS where  the  signal-to-noise is
highest.

\begin{figure}[ht]
\begin{center}
\includegraphics[width=84mm]{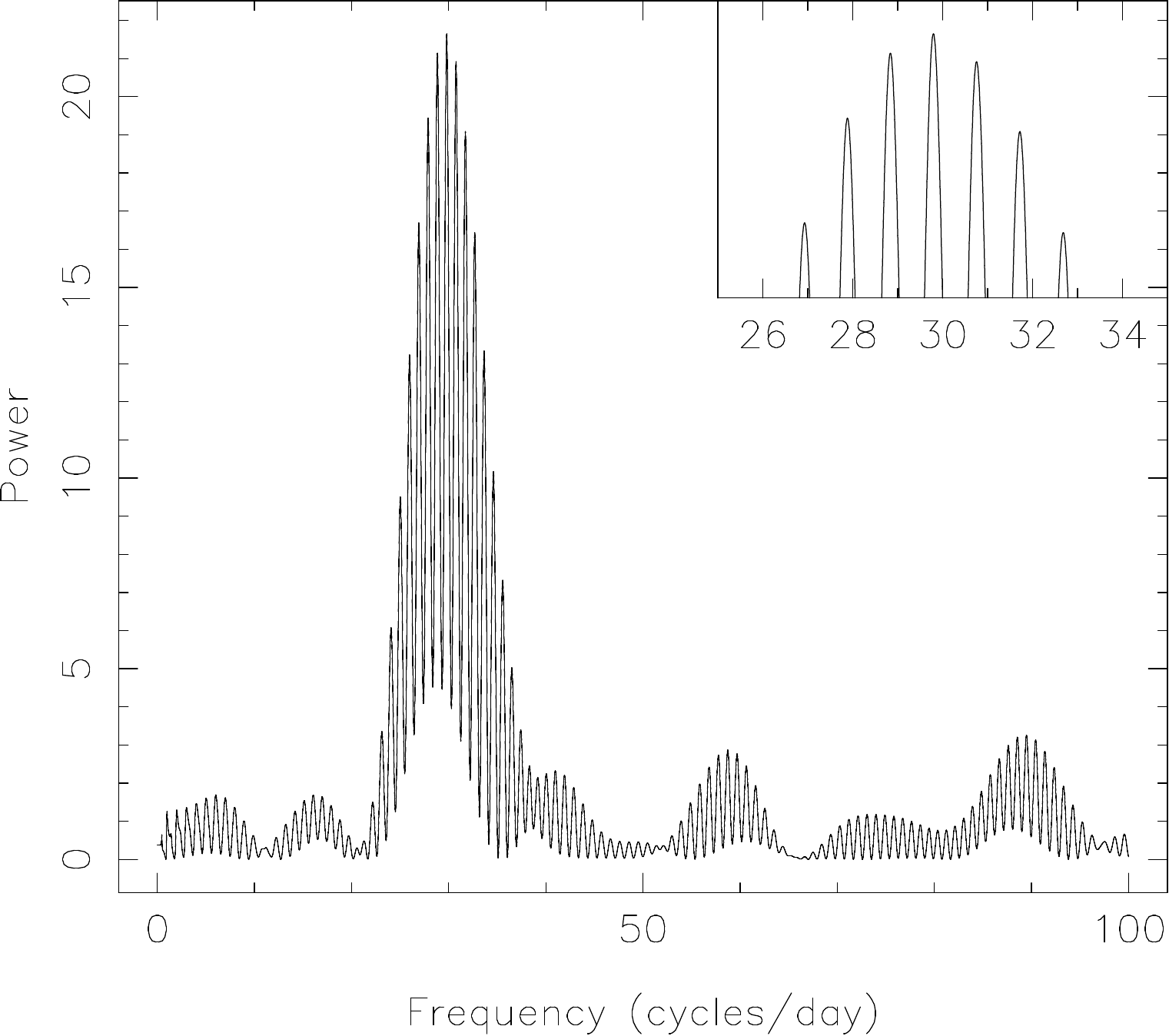}
\caption{Lomb--Sargle periodogram of the flux ratios of the blue and red wings of the combination of the strongest HeI ($\lambda\lambda3888, 4026,4471,4921,5015$) emission lines in the blue side of the Keck-I/LRIS spectra of \SDSSs. A zoom around the peak (inset) shows the typical aliases at $\pm$1 cycle per day.}
\label{fig:tsa}
\end{center}
\end{figure}

Roughly following the approach  of \cite{Nather:1981aa} we divided the
combined lines into  red and blue wings and  calculated the ratio of
the fluxes in both wings for each individual spectrum.  A Lomb-Scargle
periodogram of the flux  ratios is shown in Figure~\ref{fig:tsa}.  The
maximum  power is found  at 29.8  cylces per  day, corresponding  to a
period  of 48.3\,min.  The  periodogram shows  the next  two strongest
peaks  at  the  usual $\pm1$  cycle  per  day  aliases (see  inset  of
Figure~\ref{fig:tsa}),  which  are  caused  by  the  relatively  short
observing baseline on two consecutive nights.

\begin{figure}[ht]
\begin{center}
\includegraphics[width=84mm]{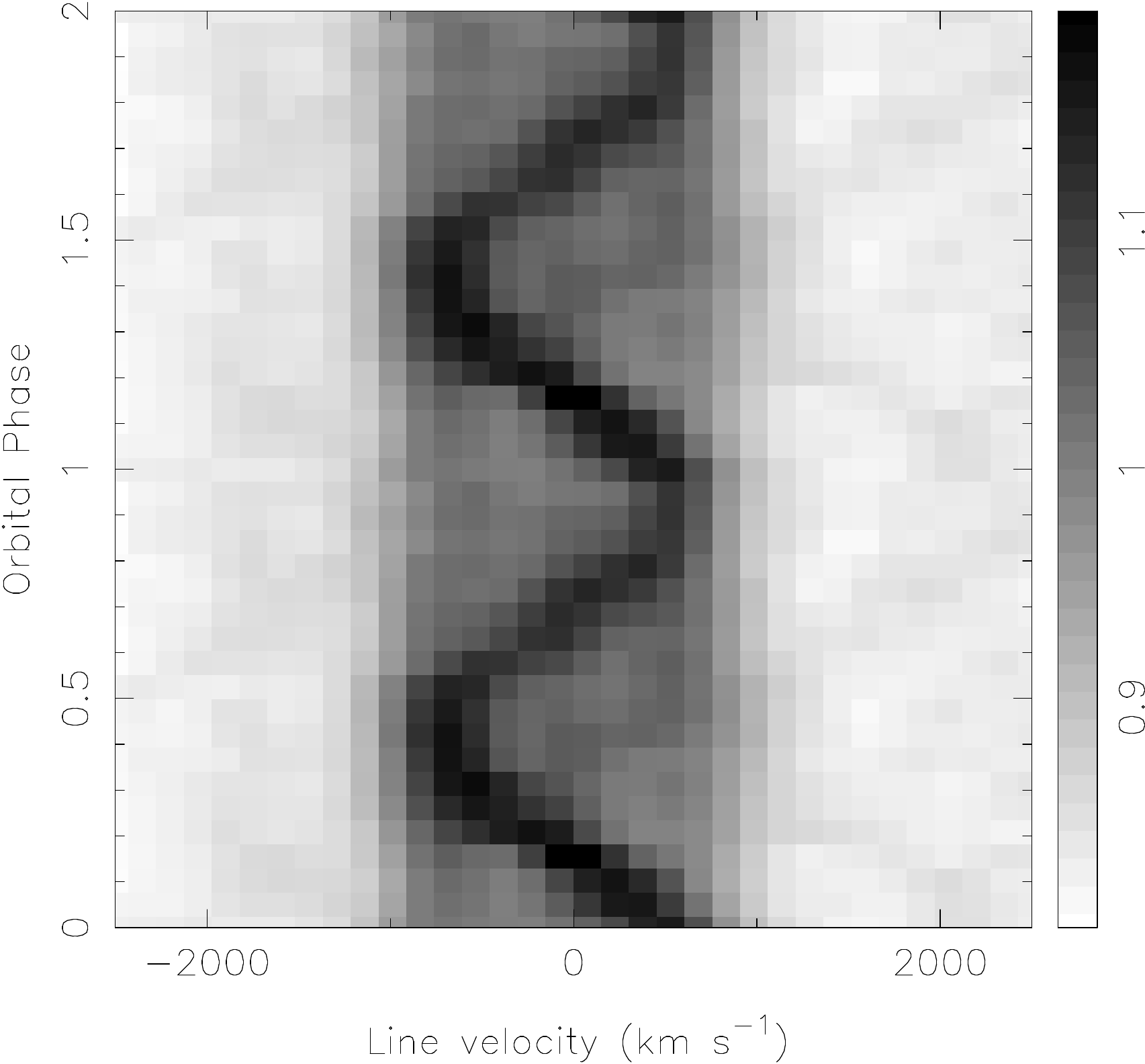}
\caption{Trailed spectrum of the combined He emission lines in the blue arm spectra of \SDSSs, phase-folded on the 48.31\,min period. The zero-phase is arbitrary. The grey-scale indicates the relative flux densities. An arbitrary zero-phase of HJD=2454858.049150 has been used.}
\label{fig:trail}
\end{center}
\end{figure}

\begin{figure}[ht]
\begin{center}
\includegraphics[width=84mm]{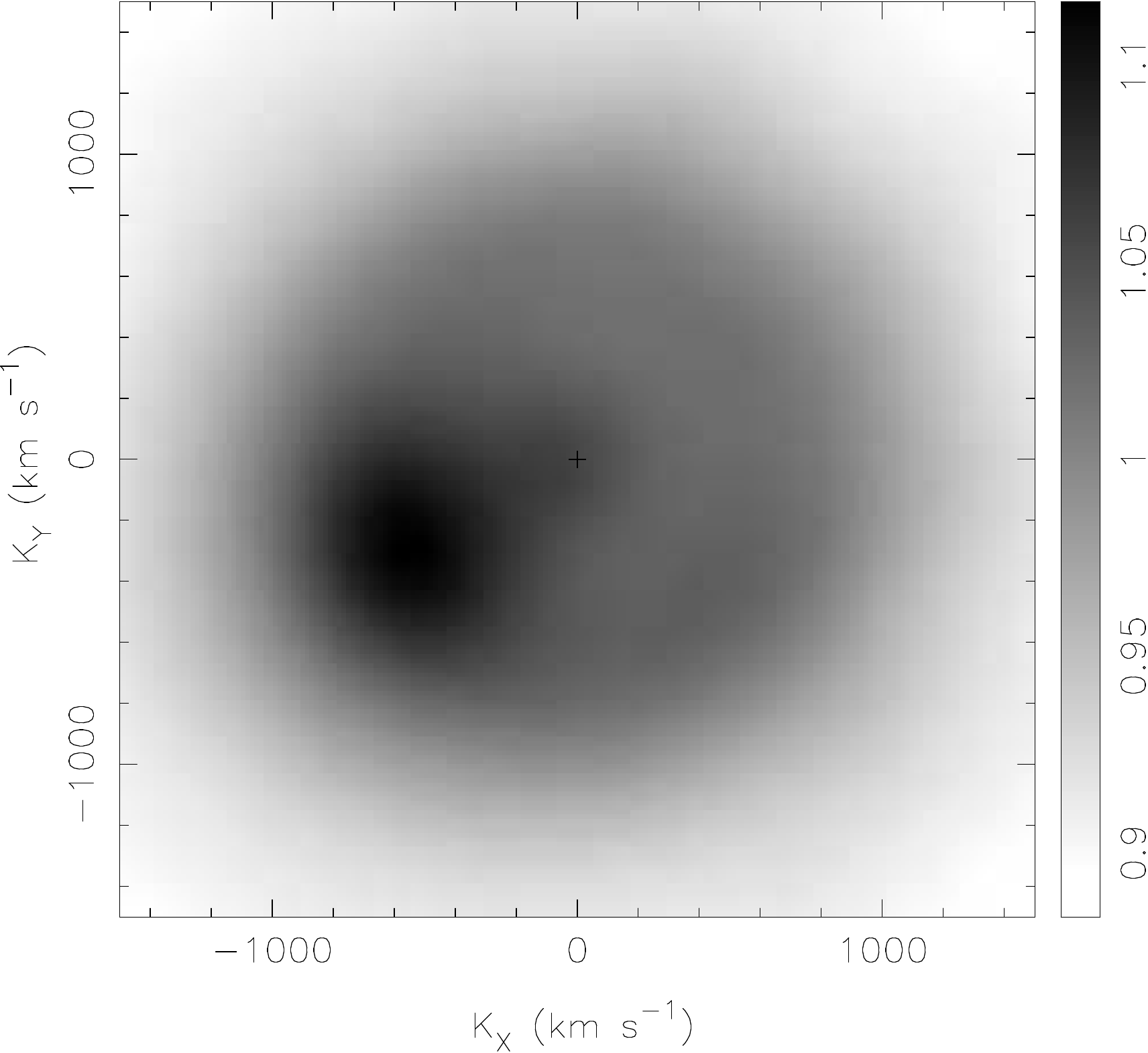}
\caption{Linear back-projection Doppler tomogram of the combined \SDSSs\ He emission lines in the blue arm spectra taken with Keck-I/LRIS, projected on the 48.31\,min period. The grey-scale indicates the relative flux densities. Zero-phase as in Figure~\ref{fig:trail}.}
\label{fig:dopplermap}
\end{center}
\end{figure}

Phase-folding the  individual spectra  of both nights  on a  period of
48.3\,min     produces    the     trailed     spectrum    shown     in
Figure~\ref{fig:trail}.   A clear  sinusoidal S-wave  with  a velocity
amplitude  of  $\approx700$\,km  s$^{-1}$  is  found.   The  observing
baseline  on the  second  night (see  Table~\ref{tab:obslog}) is  long
enough and  the S-wave signal  is strong enough  that we can  obtain a
unique trace of the S-wave signal throughout the two-night baseline of
our  Keck  run,  without  losing  count of  the  orbital  cycles.   To
accurately track the phase of the S-wave signal, we back-projected the
spectra into a Doppler tomogram \citep{Marsh:1988aa}, transforming the
S-wave   emission   into    a   localized   emission   `bright   spot'
(Figure~\ref{fig:dopplermap}).    Fine-tuning  of   the  phase-folding
period  to eliminate  the  phase drift  of  the bright  spot gives  an
orbital  period $P_\mathrm{orb}=48.31\pm0.08$  minutes.  The  error on
the orbital  period is  obtained from an  estimate of  the bright-spot
phasing  accuracy  due  to   statistical  and  systematic  errors,  as
discussed in \cite{Roelofs:2007ad}.

\subsection{Average Spectrum of \SDSSs}

\begin{figure*}[t]
\begin{center}
\includegraphics[width=0.44\textwidth,angle=90]{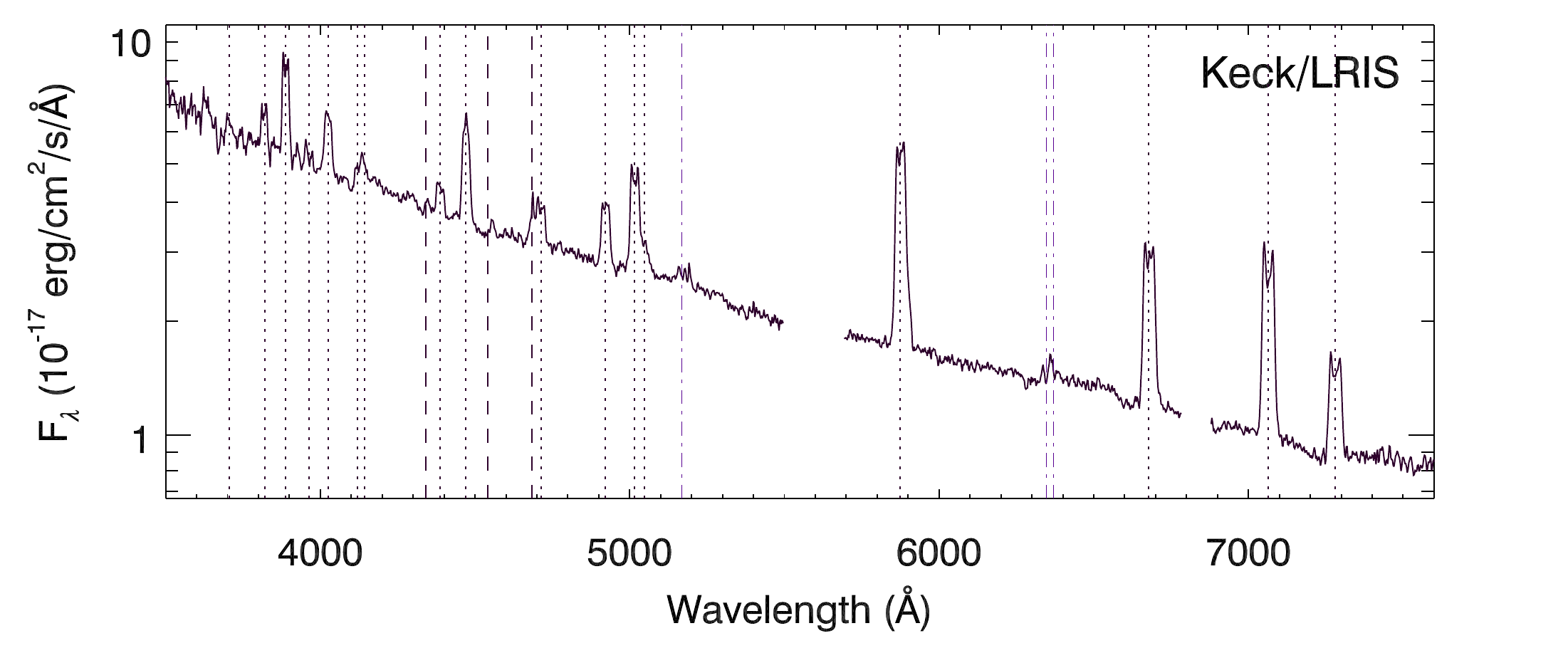}
\caption{Average Keck-I/LRIS spectrum of \SDSSs. Dotted and dashed lines mark the \ion{He}{1} and \ion{He}{2} emission features, respectively. The locations of \ion{Fe}{2} ($\lambda5169$) and \ion{Si}{2} ($\lambda6347,6371$) in emission are indicated by the dash-dotted lines.}
\label{fig:keck}
\end{center}
\end{figure*}

In  order to generate  an averaged,  high signal-to-noise  spectrum of
\SDSSs, we  combined 
all LRIS exposures  using optimal weights (Figure~\ref{fig:keck}).  As
already  indicated by  the DBSP  spectrum (Figure~\ref{fig:spectra}a),
\SDSSs\  exhibits a  blue continuum  dominated by  strong  emission of
neutral  helium, typical  for  AM~CVn stars.   In  addition, the  LRIS
spectrum reveals singly ionized helium ($\lambda4685$) as well as weak
components    of    \ion{Fe}{2}    $\lambda5169$    and    \ion{Si}{2}
$\lambda6347+6371$.  These  features are  commonly observed in  AM CVn
stars  \citep[e.g.][]{Groot:2001aa,Roelofs:2006ab},  and are  expected
from   a  donor   star   with  solar   abundances   of  heavy   metals
\citep{Marsh:1991aa}.   We note  the absence  of significant  DB white
dwarf       absorption       \citep[as       seen       in       e.g.,
SDSS~J1240-0159,][]{Roelofs:2005aa}  underlying  the  helium  emission
lines.   There are  two  known AM~CVn  stars  with comparable  orbital
period -  GP~Com \citep[P$_{\rm orb}=46.5$\,min;][]{Nather:1981aa} and
SDSS~J1411+4812                                          \citep[P$_{\rm
  orb}=46\pm2$\,min;][]{Roelofs:2007}. Overall the spectrum of \SDSSs\
is    particularly   reminiscent    of    that   of    SDSS~J1411+4812
\citep{Anderson:2005aa}.   It shows  similar metal  lines and  He line
equivalent    widths    \citep[e.g,    \ion{He}{1}    5876    EW$_{\rm
  J0902}=79\pm3$\,\AA\  (see Table~\ref{tab:sdss_phot})  vs.  EW$_{\rm
  J1411}\approx65$\,\AA;][]{Anderson:2005aa}.   In contrast,  the high
abundance of nitrogen  in GP~Com is not seen  in \SDSSs, which suggest
that the CNO cycle is less prominent in the latter.


\begin{figure}[t]
\begin{center}
\includegraphics[width=0.38\textwidth,angle=90]{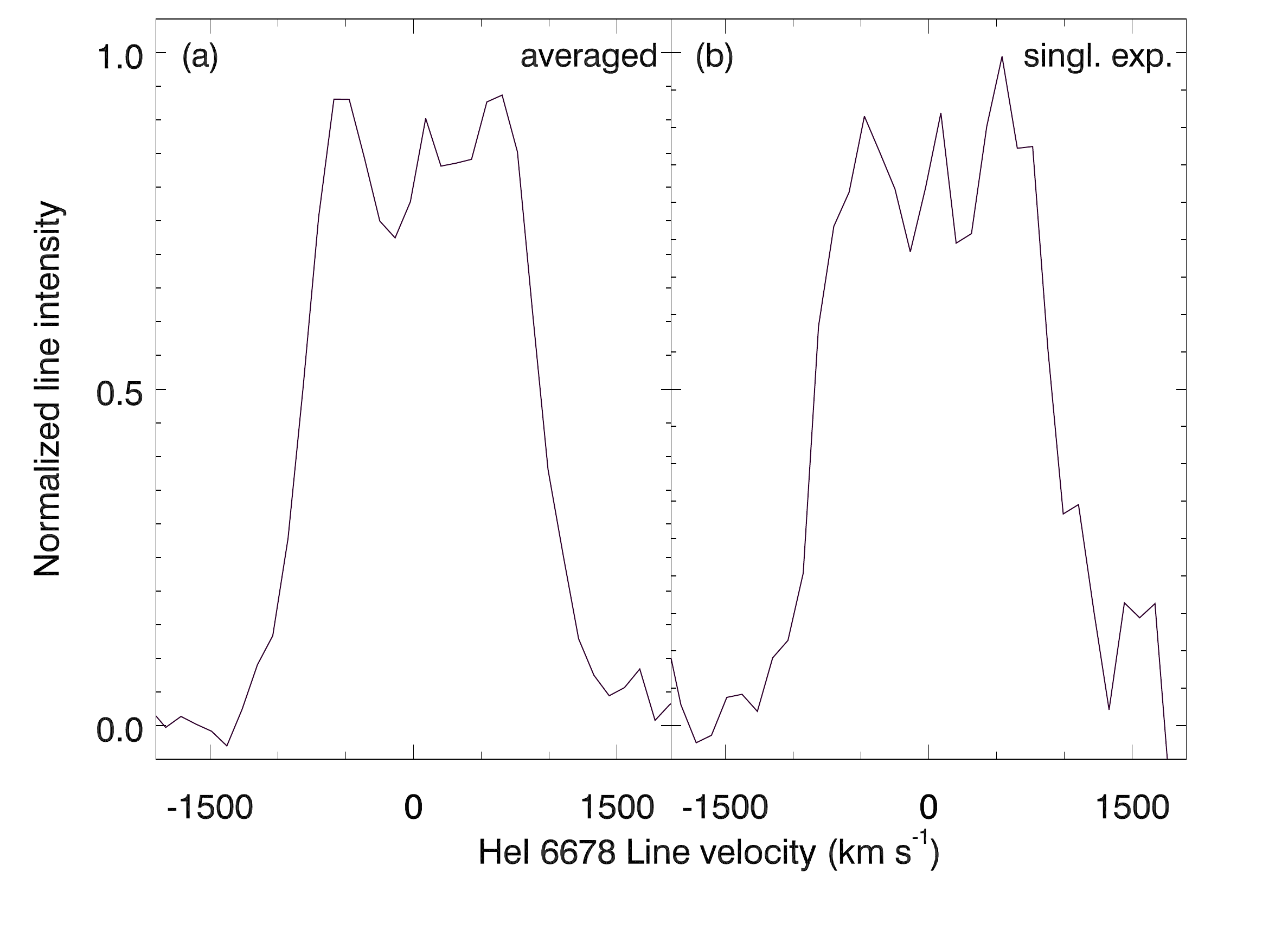}
\caption{Profile of the HeI 6678 emission line in (a) the averaged spectrum of \SDSSs\ and (b) in a single 180\,s exposure.}
\label{fig:lines}
\end{center}
\end{figure}

The  \ion{He}{1}  emission  lines,  averaged  as  well  as  in  single
exposures, show  the classical double-winged profile  indicative of an
accretion  disk (Figure~\ref{fig:lines}).   In addition,  we  detect a
third emission component  near zero velocity, similar to  that seen in
many other AM~CVn stars.  This  emission spike is thought to originate
from  (or very  near) the  surface of  the accreting  white  dwarf, as
suggested  by  observed radial  velocity  shifts  consistent with  the
accretor's      expected       orbital      motion      in      GP~Com
\citep{Marsh:1999aa,Morales-Rueda:2003aa}.  It  is interesting to note
that  the \ion{He}{2}  emission  line at  $\lambda4685$ (blended  with
\ion{He}{1}  $\lambda4713$)  has  a   Gaussian  profile  and  a  lower
full-width  at half-maximum  (FWHM, $\approx  700$\,km\,s$^{-1}$) than
the \ion{He}{1} lines  ($\approx 1200$\,km\,s$^{-1}$), suggesting that
the  \ion{He}{2}  line  may  largely  consist  of  the  central  spike
component.

\begin{figure*}[t]
\begin{center}
\includegraphics[width=0.36\textwidth,angle=90]{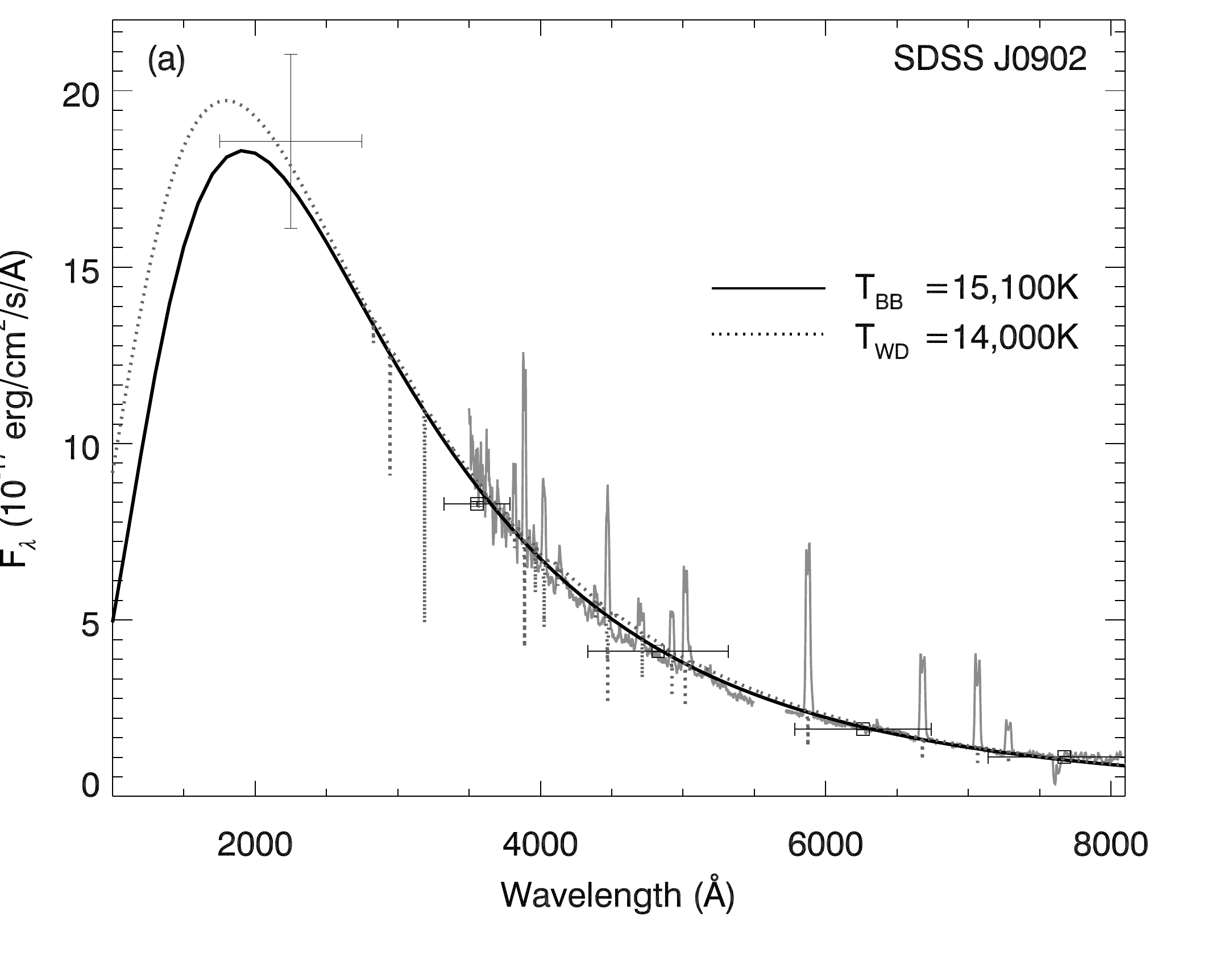}
\includegraphics[width=0.36\textwidth,angle=90]{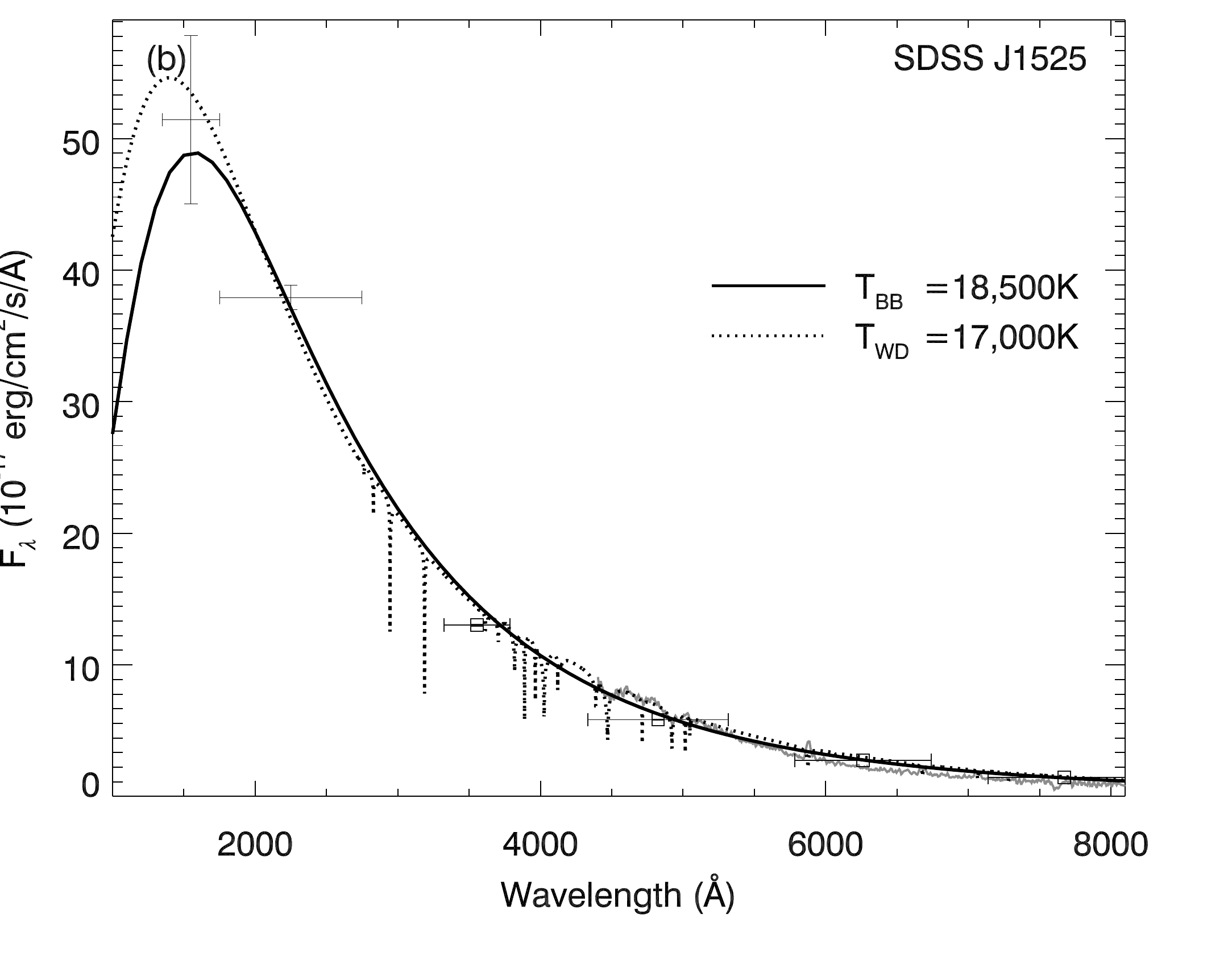}
\includegraphics[width=0.36\textwidth,angle=90]{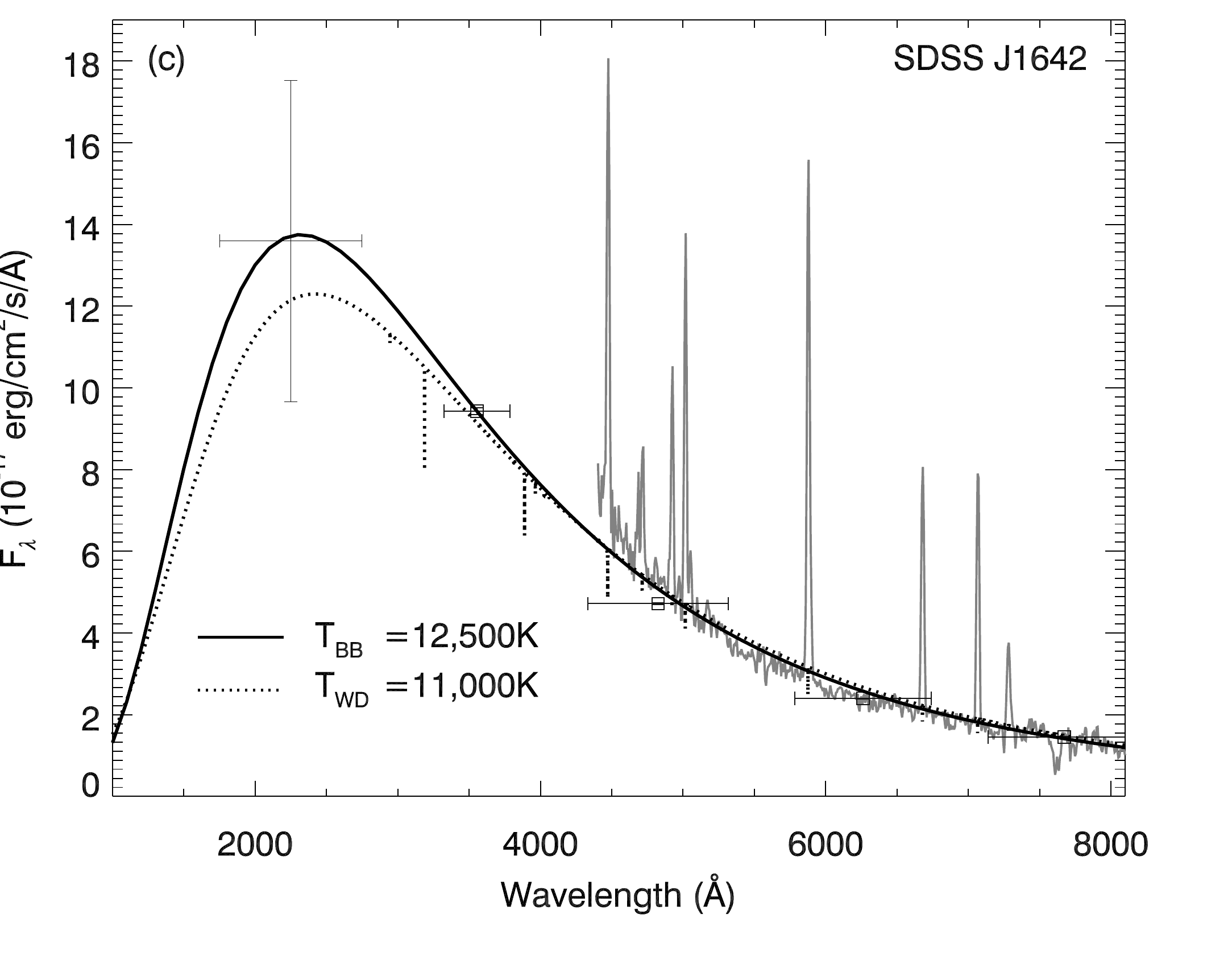}
\includegraphics[width=0.36\textwidth,angle=90]{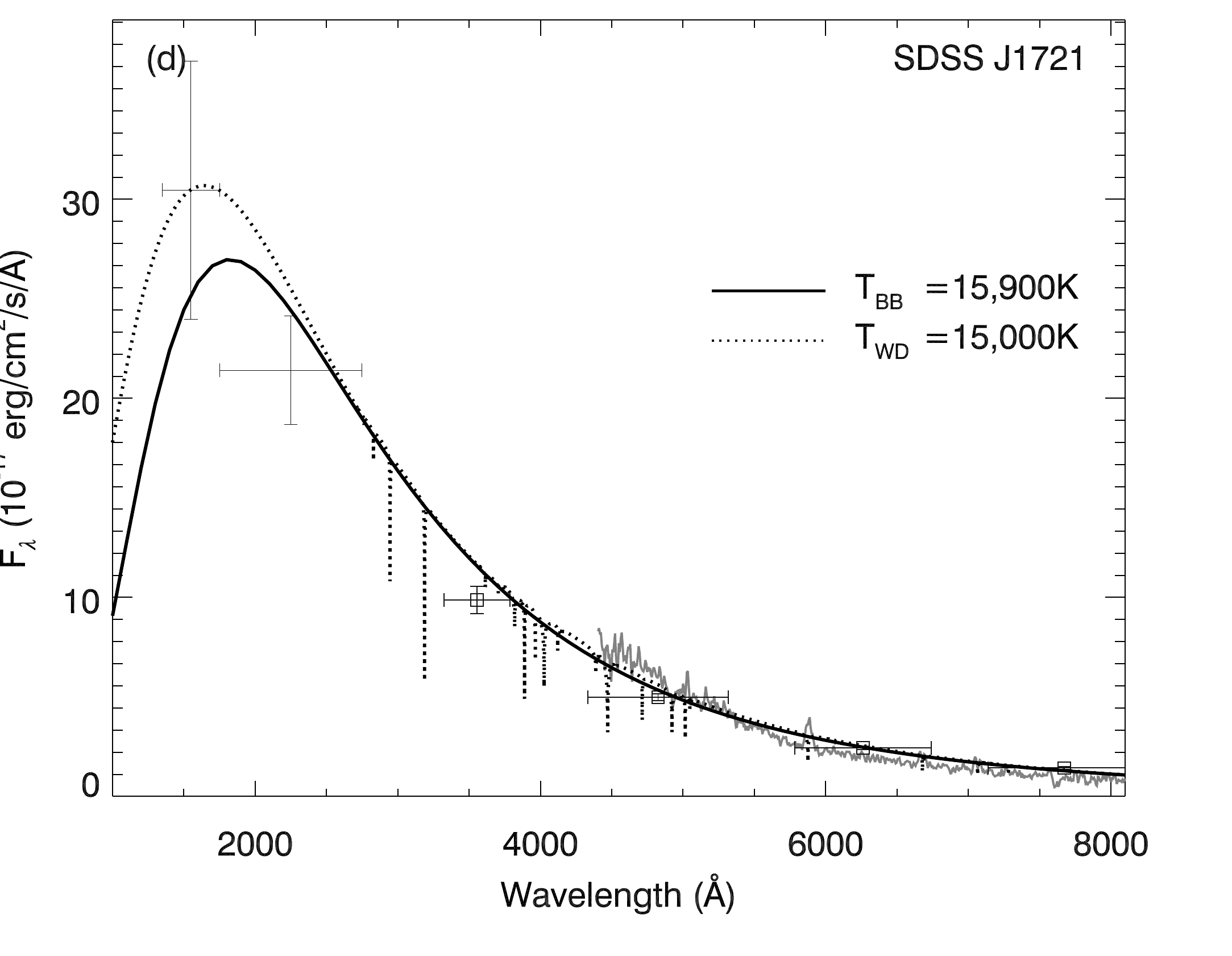}
\caption{Flux-calibrated spectra  together with  photometry from GALEX GR5 (crosses) and SDSS DR7 (empty squares, see Table~\ref{tab:sdss_phot}) for the four new AM CVn candidates. The photometry and specroscopy have been corrected for Galactic foreground extinction. The solid and dotted lines indicate black bodies and helium white dwarf atmosphere models of \cite{Koester:2008aa} with surface gravity log $g=8$, respectively.  The effective temperatures are noted in the panels. Spectrosopic observations of \SDSSs\ (panel a), SDS~J1525 (b), and SDSS~J1721 (d) have been obtained under non-photometric conditions resulting in considerable uncertainties in the absolute flux calibration. Thus, the spectra have been scaled to the SDSS photometry for visualiziation purpose.}
\label{fig:sed}
\end{center}
\end{figure*}

An estimate of the continuum temperature of \SDSSs\ can be obtained by
modeling the  available photometric and  spectroscopic data. A  fit to
the  SDSS photometry  taken  in  2002 and  the  GALEX NUV  measurement
obtained in  2004 suggest a  black body with effective  temperature of
$T_{\rm eff}\approx15,100$\,K (Figure~\ref{fig:sed}a). A similarly good
match is  obtained using the  helium white dwarf atmosphere  models of
\cite{Koester:2008aa} with $T_{\rm eff}\approx14,000$\,K.

At  the  orbital  period  of  \SDSSs, the  accreting  white  dwarf  is
expected to  contribute  a  large  fraction  ($\approx50$\,\%)  of  the
continuum  emission, if the  mass transfer  rate is  not significantly
higher than  the equilibrium rate set  by gravitational-wave radiation
\citep[$\approx10^{-11}$\,\msun\  yr$^{-1}$;][]{Bildsten:2006aa}.  This
means that the accreting dwarf is expected to be the hottest component
in the system.  Interestingly, our temperature estimate is at the high
end of expectations from  theoretical models for the thermal evolution
of  the accreting  white dwarf  \citep{Bildsten:2006aa}.  These models
predict  temperatures  in the  range  of  9,000\,K  to 15,000\,K  (for
primary  masses from 0.65\,\msun\  to 1.05\,\msun)  at orbital  periods of
$\approx48$\,min.

Since a higher system mass leads to more accretion heating and shorter
cooling time  for the  accretor at a  give orbital period,  the rather
high continuum  temperature in  \SDSSs\ may point  to a  high accretor
and/or donor mass.  The absence  of broad DB absorption lines, seen in
other   AM~CVn   stars  \citep[e.g.\   SDSS\,J1240;][]{Roelofs:2006aa}
suggests that  the accretor cannot  be much hotter than  15,000\,K, if it
contributes substantially to the optical flux as expected.


\begin{table*}[ht]
\begin{center}
\caption{GALEX GR5 and SDSS DR7 photometry and HeI$\lambda$5876 equivalent width of the new AM CVn stars.
\label{tab:sdss_phot}}
\begin{tabular}{l l l l l l l l l c}
\hline\hline
SDSS\,\ldots  & FUV$_{\rm AB}$  & NUV$_{\rm AB}$ & u & g  &  r & i & z & A(g) & --EW$_{\rm HeI\lambda5876}$ (\AA)\\
\hline
J090221.35+381941.9~ & \tablenotemark{a}20.61(29) & 20.35(17) & 20.16(5) & 20.23(2) & 20.47(3) & 20.59(5) & 20.79(21) & 0.09 & 79(3)\\
J152509.57+360054.5~ & 20.01(15) & 19.53(3) & 19.64(3) & 19.82(2) & 20.06(2) & 20.31(3) & 20.40(12) & 0.06 & 20(3)\\
J164228.06+193410.0~ &  \ldots & 21.11(31) & 20.29(5) & 20.28(2) & 20.37(2) & 20.42(3) & 20.61(19) & 0.29 & 105(3)\\
J172102.48+273301.2~ & 20.67(23) & 20.26(14) & 20.02(4) & 20.06(2) & 20.23(2) & 20.35(3) & 20.25(11) & 0.12 & 30(6)\\
\hline
\end{tabular}
\tablecomments{Photometry and g-band extinction are given in magnitudes. Numbers in parentheses indicate the errors in the corresponding number of last digits.}
\tablenotetext{a}{Visual inspection of the GALEX All-sky Imaging Survey image AIS\_85\_sg82 puts doubts on the GR5 catalogued detection, as confirmed by the non-detection in a second visit (AIS\_85\_sg72).}
\end{center}
\end{table*}


\subsection{SDSS~J1525, SDSS~J1642, and SDSS~J1721}

The  two objects  with  the weakest  emission  lines, SDSS\,J1525  and
SDSS\,J1721,  turn over to  a helium  absorption-line spectrum  in the
blue, quite  like the first Sloan-discovered AM  CVn star, SDSS\,J1240
\citep{Roelofs:2005aa}.  Black body  and helium white dwarf atmosphere
model     fits    to     their    SDSS     and     GALEX    photometry
(Figure~\ref{fig:sed}b,d)    suggest     temperatures    of    T$_{\rm
  eff}=17,000-18,500$\,K   and   T$_{\rm  eff}=15,000-15,900$\,K   for
SDSS~J1525 and SDSS~J1721, respectively. This, together with the broad
helium absorption lines in the spectra indicates orbital periods below
$\approx40$\,min    \citep{Nelemans:2005lr}.   The    relatively   low
temperature of the latter  surprises given that similarly broad helium
absorption  lines in both  systems (see  Figure~\ref{fig:spectra}). We
matched the white dwarf atmosphere models to the spectral range aodunr
the 4921\,\AA\  absorption line  and found a  best fit  temperature of
18,000\,K  in both  sources. The  apparent discrepancy  for SDSS~J1721
suggests  that   another  component  e.g.,   the  accretion  disk,
contributes signifcantly to the continuum emission.

The third source, SDSS~J1642,  shows remarkably strong helium emission
lines,  comparable to  CE~315,  the system  with  the longest  orbital
period  \citep[$65.1\pm0.7$\,min;][]{Ruiz:2001aa}.    The  absence  of
helium   absorption    and   the   low    effective   temperature   of
$\approx12,000$\,K (Figure~\ref{fig:sed}c) suggest that SDSS~J1642 may
be of long orbital period ($>50\,\min$) where the accretor is expected
to dominate the optical flux.


\section{Conclusions}
\label{sec:conclusion}

We have  presented four proposed  new AM CVn stars,  and time-resolved
spectroscopy confirming  beyond doubt the  ultra-compact binary nature
of one of  them, \SDSS. This increases the total  count of new systems
to five, after completing  the spectroscopic follow-up fo approx. half
of  the 1500 candidates  in our  sample. Based  on the  present sample
completeness (Figure~\ref{fig:sample})  we expect  close to 8  more AM
CVn  stars among the  remaining candidates.  This already  indicates a
lower space density than predicted earlier \citep{Roelofs:2007aa}, btu
a  detailed population  study will  be presented  in a  separate paper
after the follow-up program has been completed.

\SDSSs\  is the  second confirmed  AM CVn  star found  in  our ongoing
spectroscopic  survey  of  color-selected  candidates from  the  Sloan
Digital Sky  Survey, after SDSS~J0804  \citep{Roelofs:2009aa}, and has
an  orbital  period   of  $P_{\rm  orb}=48.31\pm0.08$\,min.   With  an
apparent   temperature    of   $T_\mathrm{eff}\approx$14,000\,K,   its
optical/UV  spectrum   is  at  the  high  end   of  expectations  from
theoretical   cooling  models,   assuming   gravitational-wave  driven
evolution.  This  is becoming a  common observational feature  and may
suggest large  component masses in  many AM CVns, or  angular momentum
losses additional to gravitational radiation.

At g=20.23\,mag$_{\rm AB}$, \SDSSs\  is near the faint magnitude limit
of  our sample.   Our successful  follow-up shows  that phase-resolved
spectroscopy  of  faint systems  can  be  obtained  at 8--10\,m  class
telescopes within a modest amount of time, thanks to the strong S-wave
signals  typical   of  the   emission-line  AM  CVn   stars.   Similar
observations  of the other  new candidates  are encouraged  to confirm
their  ultra-compact binary  nature and  build up  the  orbital period
distribution,  which  holds valuable  information  about the  Galactic
population  as  a  whole  and  the  physics  governing  its  evolution
\citep[e.g.][]{Deloye:2007aa}.


\acknowledgments 

AR  acknowledges  support  through  NASA grant  NNX08AK66G.   GHAR  is
supported  by NWO Rubicon  grant 680.50.0610.  DS acknowledges  a STFC
Advanced Fellowship. This paper is based on data obtained at the W.~M.
Keck Observatory, which is  operated as a scientific partnership among
the California Institute of  Technology, the University of California,
and  the National  Aeronautics and  Space Administration  (NASA).  The
Observatory was made possible by the generous financial support of the
W.M.   Keck  Foundation.   We  also  acknowledge use  of  the  Palomar
Hale-5\,m telescope operated by the California Institute of Technology
and Palomar Observatory; the  Nordic Optical Telescope and the William
Herschel Telescope, La Palma; and  the {\it GALEX} public archive.  We
are grateful to D.~Koester for kindly making available his white dwarf
atmosphere models.





\clearpage

\end{document}